\documentstyle{amsppt}\TagsOnRight\nologo
\newcount\refcount
\advance\refcount 1
\def\newref#1{\xdef#1{\the\refcount}\advance\refcount 1}
\newref\bennettetal
\newref\definitions
\newref\VPbound
\newref\epp
\newref\OhyaPetz
\newref\Peres
\newref\boundentanglement
\newref\KKT
\newref\werner
\newref\puredistill
\def\Tr{\operatorname{Tr}}
\def\ket#1{{|#1\rangle}}
\def\bra#1{{\langle#1|}}
\let\cal=\Cal

\def\defeq{\overset\text{def}\to=}
\topmatter
\title An improved bound on distillable entanglement \endtitle
\author 
E. M. Rains
\endauthor
\affil AT\&T Research \endaffil
\address AT\&T Research, Room C290, 180 Park Ave.,
         Florham Park, NJ 07932-0971, USA \endaddress
\email rains\@research.att.com \endemail
\date October 30, 2000\enddate
\abstract
The best bound known on 2-locally distillable entanglement is that of
Vedral and Plenio, involving a certain measure of entanglement based on
relative entropy.  It turns out that a related argument can be used to give
an even stronger bound; we give this bound, and examine some of its
properties.  In particular, and in contrast to the earlier bounds, the new
bound is {\it not} additive in general.  We give an example of a state for
which the bound fails to be additive, as well as a number of states for
which the bound {\it is} additive.

\endabstract
\endtopmatter
One of the central problems in quantum information theory is that of
entanglement distillation \cite\bennettetal, the production of maximally
entangled states from partially entangled states.  The purpose of the
present note is to give a new upper bound on the rate at which entanglement
can be distilled.

In general, if $C$ is a given class of physical operations, we define
$C$-distillable entanglement as follows:

\proclaim{Definition}
The $C$-distillable entanglement of a state $\rho$ on a state
space $V_A\otimes V_B$ is the maximum number $D_C(\rho)$ such that
there exists a sequence 
$$
{\cal T}_i:(V_A\otimes V_B)^{\otimes n_i}\to V_i\otimes V_i
$$
of operations from $C$, with $n_i\to\infty$,
$$
{1\over n_i} \log_2 \dim V_i\to D_C(\rho),
$$
and
$$
F({\cal T}_i(\rho^{\otimes n}))\to 1.
$$
\endproclaim

See \cite\definitions\ for a discussion of why this is a valid definition of
distillable entanglement.  Here $F(\sigma)$ is the {\it fidelity} of the
state $\sigma$, defined by
$$
\align
F(\sigma) &= \Phi^+(V)^\dagger \sigma \Phi^+(V),\\
\Phi^+(V) &= {1\over \sqrt{\dim V}}\sum_{1\le i\le\dim V} \ket{ii}.
\endalign
$$
Here $\Phi^+(V)$ is a chosen maximally entangled state.

The history of bounds on 2-locally distillable entanglement (for which $C$
is generated by local operations and two-way classical communication)
involves a rather curious phenomenon.  The first bound, entanglement of
formation, is most easily proved for 2-local operations.  However, the
second, stronger, bound of Vedral and Plenio (\cite\VPbound, see also
\cite\epp) applies to a {\it larger} class of operations, namely that of
separable operations.  The bound of the present note carries this even
further, both strengthening the bound and enlarging the set of allowed
operations.  This phenomenon is rather counter-intuitive, since enlarging
the class of allowed operations would be expected to increase the
distillable entanglement.

\head The new bound\endhead

As in \cite\epp, the key to the new bound is the observation that if we
are given a process that distills entanglement from $\rho$ at a given rate
and high fidelity, and apply the process to a different state $\sigma$,
then there is a limit to how much the fidelity can be reduced by doing so.
Given an upper bound on the fidelity any process of that rate can obtain
from $\sigma$, we may be able to deduce that no process can obtain high
fidelity from $\rho$.

We will state this in some generality, to support possible future
applications.

\proclaim{Theorem 1}
Let $C$ be some class of operations.  Suppose we are given a state $\sigma$
on $V_A\otimes V_B$ and an increasing, left continuous, function
$\alpha:{\Bbb R}^+\to {\Bbb R}^+$ such that any operation
$$
{\cal T}:(V_A\otimes V_B)^{\otimes n}\to V\otimes V
$$
from $C$ satisfies
$$
{1\over n}\log_2 F_{\cal T}(\sigma^{\otimes n})\le
-\alpha({1\over n}\log_2\dim V).
$$
Then for any other state $\rho$,
$$
\alpha(D_C(\rho))\le S(\rho||\sigma),
$$
where
$$
S(\rho||\sigma) = \Tr(\rho(\log_2(\rho)-\log_2(\sigma)))
$$
is the relative entropy of $\rho$ and $\sigma$.
\endproclaim

\demo{Proof}
The crucial observation is that for any operator ${\cal T}$ the function
$F_{\cal T}$ is linear on density operators, and is bounded between 0 and
1.  It follows that we can write it in the form
$$
F_{\cal T}(\omega) = \Tr(F_{\cal T} \omega)
$$
for some (uniquely determined) operator $F_{\cal T}$ such that $F_{\cal T}$
and $1-F_{\cal T}$ are both positive.

For $\epsilon>0$, let $n$, $V$, ${\cal T}$ give a
process from $C$ with
$$
{1\over n}\log_2 \dim V \ge D_C(\rho)(1-\epsilon),
$$
and
$$
\Tr (F_{{\cal T}} \rho^{\otimes n}) \ge 1-\epsilon.
$$
On the other hand, by assumption,
$$
\Tr (F_{{\cal T}} \sigma^{\otimes n})
\le 2^{-n \alpha({1\over n}\log_2\dim V)}.
$$
Lemma 1 below then tells us that
$$
\align
S(\rho||\sigma)
&\ge
\limsup_{\epsilon\to 0} \alpha({1\over n} \log_2\dim V)\\
&\ge
\limsup_{\epsilon\to 0} \alpha((1-\epsilon) D_C(\rho))\\
&=
\alpha(D_C(\rho)).
\endalign
$$
\qed\enddemo

Remark.  The above argument is a hybrid of the arguments of \cite\VPbound\
and \cite\epp.  In particular, it should be noted that the above argument
does not require any assumption that the resulting bound be additive.

\proclaim{Lemma 1}
Let $\rho$ and $\sigma$ be states on a common Hilbert space $V$, such that
$S(\rho||\sigma)$ is finite.  For $n\in {\Bbb Z}^+$ and $0<\epsilon<1$,
define
$$
R(n,\epsilon)=\inf_\pi\{\log_2 \Tr(\sigma^{\otimes n} \pi)\},
$$
where $\pi$ ranges over positive operators on $V^{\otimes n}$ with
both $\pi$ and $1-\pi$ positive, and such that
$$
\Tr(\rho^{\otimes n} \pi)\ge 1-\epsilon.
$$
Then
$$
\liminf_{\epsilon\to 0} \lim_{n\to\infty} {1\over n} R(n,\epsilon) \ge
-S(\rho||\sigma).
$$
\endproclaim

\demo{Proof}
From the condition on $\pi$, it follows that $\{\pi,1-\pi\}$ is a POVM.
Consequently, Uhlmann's monotonicity theorem (\cite\OhyaPetz, theorem 1.5)
tells us that, writing
$$
\align
p_\rho &= \Tr(\rho^{\otimes n} \pi),\\
p_\sigma &= \Tr(\sigma^{\otimes n} \pi),
\endalign
$$
we have
$$
n S(\rho||\sigma)
\ge
p_\rho (\log_2(p_\rho)-\log_2(p_\sigma))+
(1-p_\rho) (\log_2(1-p_\rho)-\log_2(1-p_\sigma)).
$$
Now, this in turn is bounded below by
$$
-1-
p_\rho \log_2(p_\sigma)-
(1-p_\rho) \log_2(1-p_\sigma).
$$
If we divide by $n$, the first and third terms will be negligible unless
$$
p_\sigma=\Tr(\sigma^{\otimes n}\pi)>{1\over 2},
$$
say.  But such a $\pi$ could not possibly provide a counterexample to the
lemma.  It remains to consider the second term.
But
$$
-{1\over n}p_\rho\log_2(p_\sigma)
\ge
-(1-\epsilon) {1\over n}\log_2(p_\sigma).
$$
It follows that
$$
-(1-\epsilon) \lim_{n\to\infty} {1\over n}\log_2(p_\sigma)
\le
S(\rho||\sigma).
$$
The lemma follows by taking the limit as $\epsilon\to 0$.
\qed\enddemo

Remark.  Indeed, it is the case that
$$
\lim_{\epsilon\to 0} \lim_{n\to\infty} {1\over n} R(n,\epsilon)
=
-S(\rho||\sigma)
$$
(\cite\OhyaPetz, equation 1.31), in which the operator $\pi$ may be assumed
to be a projection.  In particular the conclusion of theorem 1 is the
strongest that can be made from its hypotheses.

To obtain a bound, then, we need to find states $\sigma$ for which
we can bound the fidelity.  Let $\Gamma$ denote the partial transpose
operator of \cite\Peres.  Let $C_\Gamma$ be the set of
positive-partial-transpose (p.p.t.) superoperators, that is
completely positive, trace-preserving superoperators ${\cal S}$
such that the superoperator
$$
{\cal S}^\Gamma: \rho\mapsto ({\cal S}(\rho^\Gamma))^\Gamma
$$
is also completely positive.  It is not too difficult to see that any
separable superoperator is also p.p.t., and thus any 2-local superoperator
is p.p.t.  Similarly, we will say a state $\rho$ is p.p.t.  if
$\rho^\Gamma$ is positive semi-definite.  Note that the creation of a
p.p.t. state is a p.p.t. operation.  We will write ``$\Gamma$-distillable
entanglement'' for ``$C_\Gamma$-distillable entanglement''.

The key observation \cite\boundentanglement\ is that p.p.t. states must
have a $\Gamma$-distillable entanglement of 0, because the fidelity of a
p.p.t. state can be bounded away from 1.  Indeed:

\proclaim{Lemma 2}
Let $\sigma$ be a p.p.t. state on a space $V\otimes V$.
Then
$$
F(\sigma)\le {1\over \dim V}.
$$
\endproclaim

\demo{Proof}
We have
$$
\align
F(\sigma)
&= \Tr(\Phi^+(V) \Phi^+(V)^\dagger \sigma)\\
&= \Tr((\Phi^+(V) \Phi^+(V)^\dagger)^\Gamma \sigma^\Gamma).
\endalign
$$
But 
$$
(\Phi^+(V) \Phi^+(V)^\dagger)^\Gamma
=
{1\over \dim V}
\sum_{i,j} \ket{ij}\bra{ji},
$$
and thus has eigenvalues of absolute value at most $1/\dim V$.  So
$$
\Tr((\Phi^+(V) \Phi^+(V)^\dagger)^\Gamma \sigma^\Gamma)
\le
{1\over \dim V}\Tr(\sigma^\Gamma)
=
{1\over \dim V}.
$$
\qed\enddemo

Remark.  The new bound is essentially an update of the Vedral-Plenio bound
to take this observation into account.

\proclaim{Theorem 2}
For any state $\rho$ and any p.p.t. state $\sigma$ on the same bipartite
Hilbert space,
$$
D_\Gamma(\rho)\le S(\rho||\sigma).
$$
\endproclaim

\demo{Proof}
It suffices to show that the function $\alpha$ of theorem 1 can be taken
to be 1.  In other words, we must show that for any p.p.t. superoperator
$$
{\cal T}:(V_A\otimes V_B)^{\otimes n}\to V\otimes V,
$$
we have
$$
F_{\cal T}(\sigma^{\otimes n})\le {1\over \dim V}.
$$
But the image of a p.p.t. operator under a p.p.t. superoperator is p.p.t.,
since
$$
{\cal T}(\omega)^\Gamma
=
{\cal T}^\Gamma(\omega^\Gamma),
$$
so the lemma applies.
\qed\enddemo

The statement that
$$
F_{\cal T}(\sigma^{\otimes n})\le {1\over \dim V}
$$
for a p.p.t. superoperator ${\cal T}$ and a p.p.t. state $\sigma$ is
a special case of the following:

\proclaim{Theorem 3}
Let ${\cal T}$ be a p.p.t. superoperator with output dimension $K$
and associated fidelity operator $F_{\cal T}$.  Then
$$
-{1\over K}\le F^\Gamma_{\cal T} \le {1\over K},
$$
where for Hermitian operators $A$ and $B$, $A\le B$ means that $B-A$ is
positive semi-definite.
\endproclaim

\demo{Proof}
It suffices to show that for any density operator $\rho$,
$$
-{1\over K}\le \Tr(F^\Gamma_{\cal T}\rho)\le {1\over K}.
$$
But
$$
\align
\Tr(F^\Gamma_{\cal T} \rho) &=
\Tr(F_{\cal T} \rho^\Gamma) \\
&=
\Tr(\Phi^+(K)\Phi^+(K)^\dagger {\cal T}(\rho^\Gamma))\\
&=
\Tr(\Phi^+(K)\Phi^+(K)^\dagger ({\cal T}^\Gamma(\rho))^\Gamma)\\
&=
\Tr((\Phi^+(K)\Phi^+(K)^\dagger)^\Gamma {\cal T}^\Gamma(\rho)).
\endalign
$$
Since $-(1/K)\le (\Phi^+(K)\Phi^+(K)^\dagger)^\Gamma \le (1/K)$, and ${\cal
T}^\Gamma(\rho)$ is a density operator, the result follows.
\qed\enddemo

Remarks. (1) It is an open question whether this inequality, together with the
inequality
$$
0\le F_{\cal T}\le 1
$$
which holds for all superoperators, can be used to give a stronger bound
than that of Theorem 2, which essentially only uses the inequalities
one at a time.  (2) This statement is essentially a generalization of
equation (16) of \cite\epp.

\head Optimizing $\sigma$\endhead

To obtain the full strength of the bound of theorem 2, it is necessary
to optimize the choice of $\sigma$.  In the sequel, we will say that
a p.p.t. $\sigma$ is optimal for $\rho$ if
$$
S(\rho||\sigma) = \min_{\sigma'} S(\rho||\sigma') \defeq B_\Gamma(\rho)
$$
where $\sigma'$ ranges over all p.p.t. states.

\proclaim{Theorem 4}
Suppose $\rho$ is a positive definite state.  Then the p.p.t. state
$\sigma$ is optimal for $\rho$ if and only if, setting
$$
K=1-D_\sigma \Tr(\rho\log(\sigma)),
$$
where $D_\sigma$ is the matrix derivative, we have
$$
\sigma^\Gamma K^\Gamma = 0\quad\text{and}\quad K^\Gamma>0.
$$
\endproclaim

\demo{Proof}
We note first that $\sigma$ must also be positive definite. Otherwise,
$S(\rho||\sigma)$ would be infinite, but this is impossible, since
$S(\rho||{1\over \dim\rho})<\infty$.  Thus we must solve the optimization
problem:

Minimize $S(\rho||\sigma)$ subject to the constraints $\Tr\sigma=1$,
$\sigma>0$ and $\sigma^\Gamma\ge 0$.

Now, $S(\rho||\sigma)$ is a convex function (\cite\OhyaPetz, theorem 1.4),
that is:
$$
S(\rho||a\sigma_1+(1-a)\sigma_2)
\le
a S(\rho||\sigma_1)+
(1-a) S(\rho||\sigma_2).
$$
Moreover, the set of p.p.t. density operators is convex.  Thus $\sigma$
is optimal for $\rho$ if and only if it satisfies the Karush-Kuhn-Tucker
conditions (\cite\KKT, Theorem 2.1.4).  For the set of p.p.t. density
operators, this becomes:
$$
D_\sigma \Tr(\rho\log(\sigma)) + \lambda + K=0,
$$
for some number $\lambda$ and Hermitian matrix $K$, where $K^\Gamma$ is
positive and supported on the kernel of $\sigma^\Gamma$.  Multiplying on
the left by $\sigma$ and taking a trace, we find
$$
\align
-\lambda &= \Tr(\sigma D_\sigma\Tr(\rho\log(\sigma)))\\
&={d\over dt} \Tr(\rho\log(\sigma+t\sigma))\\
&={d\over dt} \Tr(\rho\log(1+t))\\
&=1.
\endalign
$$
\qed\enddemo

If $\rho$ is only semi-definite, then $\sigma$ can be semi-definite,
and the condition is somewhat more complicated:
$$
1-D_\sigma \Tr(\rho\log\sigma) = K + L,
$$
where $\sigma L=0$, $\sigma^\Gamma K^\Gamma=0$, and both $L$ and
$K^\Gamma$ are positive semi-definite.  For simplicity, we will assume
that $\rho$ is definite in the proofs below, but in each case, the proof
can be adapted to this more complicated case.

Remark.  In the published version of this paper, the positivity condition
on $K$ was overlooked.  The resulting condition is still valid (for $\rho$
definite and not p.p.t.) when $\sigma$ is a {\it smooth} point on the
boundary of the set of p.p.t. operators; the fact that the optimal $\sigma$
must be on the boundary forces $K$ to have at least one positive
eigenvalue.  Unfortunately, a tensor product of two boundary points is
never a smooth point on the larger cone, and thus most of the additivity
results of the published version are invalid.

One consequence of theorem 4 is that for a specific $\sigma$, it is
reasonably straightforward to determine the set of $\rho$ for which it is
optimal.  For instance, if we consider the p.p.t. state
$$
\sigma = 
\pmatrix
{1\over 6}&0&0&0\\
0&{55\over 144}&{-1\over 6}&0\\
0&{-1\over 6}&{41\over 144}&0\\
0&0&0&{1\over 6}
\endpmatrix,
$$
we find that $\sigma$ is optimal for (among others) the state
$$
\rho =
\pmatrix
{1\over 12}&0&0&0\\
0&{45907\over 90000}-{7\over 150} x&{-1201\over 3750}-{49\over 3600}x&0\\
0&{-1201\over 3750}-{49\over 3600} x&{29093\over 90000}+{7\over 150}x&0\\
0&0&0&{1\over 12}
\endpmatrix,
$$
where $x=1/\ln(73/23)$.  However, $\sigma\otimes\sigma$ is {\it not}
optimal for $\rho\otimes\rho$.  It follows immediately that
$$
B_\Gamma(\rho\otimes\rho)<2 B_\Gamma(\rho).
$$
Thus the new bound is not additive.
%This is in sharp contrast to the
%previous bounds, for which numerical evidence suggests additivity.
%(In particular, J. Smolin (personal communication) reports a computation
%indicating that the Vedral-Plenio bound is additive for the above state.)
%It is worth mentioning that if one could show that the Vedral-Plenio
%bound were actually additive, this would give an example of a state
%for which the regularized entanglement of formation is
%strictly greater than distillable entanglement.  There is reason
%to believe that this would be easier to prove than additivity of
%entanglement of formation.

Since $B_\Gamma$ is certainly subadditive, we can regularize to a
stronger bound:
$$
{\tilde B}_\Gamma(\rho) \defeq \lim_{n\to\infty} {1\over n}
B_\Gamma(\rho^{\otimes n}).
$$
It is not clear how to compute this bound, however.  It turns out, however,
that there are a number of states for which the bound {\it is} additive,
and in particular $B_\Gamma = {\tilde B}_\Gamma$.  The simplest sufficient
condition seems to be

\proclaim{Theorem 5}
Let $\rho$ be a definite state such that there exists $\sigma$ optimal for
$\rho$ that commutes with $\rho$.  Suppose further that
$$
D_\sigma(\Tr(\rho\log(\sigma)))^\Gamma\ge 0.
$$
Then for any other state $\rho'$,
$$
B_\Gamma(\rho\otimes \rho')=B_\Gamma(\rho)+B_\Gamma(\rho').
$$
Similarly, if
$$
D_\sigma(\Tr(\rho\log(\sigma)))^\Gamma\ge -1,
$$
then
$$
B_\Gamma(\rho^{\otimes n})=nB_\Gamma(\rho)
$$
for all $n\ge 1$.
\endproclaim

\demo{Proof}
Let $\sigma'$ be a state optimal for $\rho'$; assume $\rho'$ is definite
(otherwise write it as a limit of definite states).  We need only show that
$\sigma\otimes \sigma'$ is optimal for $\rho\otimes \rho'$.  We can thus
apply theorem 4.  The crucial observation is that
$$
D_{\sigma\otimes\sigma'} \Tr(\rho\otimes \rho'\log(\sigma\otimes\sigma'))
=
D_\sigma(\Tr(\rho)\log(\sigma))\otimes
D_{\sigma'}(\Tr(\rho')\log(\sigma'))
$$
whenever $[\rho,\sigma]=0$.  We may simultaneously diagonalize $\rho$
and $\sigma$.  Then by basic properties of the matrix
derivative, we have
$$
D_{\sigma\otimes\sigma'} \Tr(\rho\otimes \rho'\log(\sigma\otimes\sigma'))
=
{d\over dt} \log(\sigma\otimes\sigma' + t \rho\otimes \rho')
$$
But the argument to the logarithm is block-diagonal; we can therefore
consider each block independently.  In other words, it suffices to
consider the case in which $\sigma$ and $\rho$ are scalars.  But
then
$$
{d\over dt} \log(\sigma\sigma' + t\rho\rho')
=
{\rho/\sigma} {d\over dt} \log(\sigma'+t\rho'),
$$
as desired.

Then the first condition of theorem 4 requires
$$
(\sigma\otimes\sigma')^\Gamma
(
D_\sigma(\Tr(\rho)\log(\sigma))\otimes
D_{\sigma'}(\Tr(\rho')\log(\sigma'))
-1
)^\Gamma
=0,
$$
which is straightforward to verify.  The other condition is that
$$
(D_\sigma(\Tr(\rho)\log(\sigma))\otimes
 D_{\sigma'}(\Tr(\rho')\log(\sigma'))^\Gamma <1.
$$
But this follows since
$$
\align
D_\sigma(\Tr(\rho)\log(\sigma))^\Gamma &\ge 0\\
D_{\sigma'}(\Tr(\rho')\log(\sigma'))^\Gamma &<1.
\endalign
$$

The self-additivity claim follows similarly.
\qed\enddemo

Remark.  Note that when $[\rho,\sigma]=0$,
$D_\sigma(\Tr(\rho\log\sigma))=\rho\sigma^{-1}$.

Remark.  This, again, is weaker than the published version; we thank
R. F. Werner and K. G. H. Vollbrecht for telling us of the counterexample
they found \cite\werner.

In particular, this applies when $\rho=\sigma$, so

\proclaim{Corollary 1}
If $\rho$ is p.p.t., then for any other state $\rho'$,
$$
B_\Gamma(\rho\otimes \rho')=B_\Gamma(\rho').
$$
\endproclaim

\head Exploiting symmetries\endhead

The most powerful tool for computing an optimal $\sigma$ seems to be
the following result:

\proclaim{Theorem 6}
Let $\rho$ be an arbitrary state on $V_A\otimes V_B$.  Let $G$
be a subgroup of $U(V_A)\otimes U(V_B)$ consisting of matrices
$U_A\otimes U_B$ such that
$$
(U_A\otimes U_B) \rho (U_A\otimes U_B)^\dagger = \rho.
$$
Then there exists some $\sigma$ optimal for $\rho$ with
$$
(U_A\otimes U_B) \sigma (U_A\otimes U_B)^\dagger = \sigma
$$
for all $U_A\otimes U_B\in G$.
\endproclaim

\demo{Proof}
There certainly exists some $\sigma$ optimal for $\rho$.
The point, then, is that for any $U_A$, $U_B$,
$$
(U_A\otimes U_B)\sigma (U_A\otimes U_B)^\dagger
$$
is p.p.t., and optimal for
$$
(U_A\otimes U_B)\rho (U_A\otimes U_B)^\dagger.
$$
In particular, this is true for $U_A\otimes U_B\in G$.  But
$G$ is a closed subgroup of a compact Lie group, so there is
a unique invariant probability measure on $G$.  Define
$$
\sigma' = E_{U_A\otimes U_B\in G} (U_A\otimes U_B)\sigma (U_A\otimes
U_B)^\dagger.
$$
By convexity, $\sigma'$ is p.p.t., and
$$
S(\rho||\sigma') \le S(\rho||\sigma).
$$
Since $\sigma$ is optimal for $\rho$, so is $\sigma'$.  Since
$\sigma'$ is clearly preserved by $G$, we are done.
\qed\enddemo

Example 1.  Let $\rho$ be a state of the form
$$
a\Phi^+(V)\Phi^+(V)^\dagger + b,
$$
with $\rho^\Gamma>0$ (an {\it isotropic} state).  Then we may take $G$ to
be the group
$$
\{ U\otimes \overline{U}:U\in U(V)\}.
$$
But, in fact, $G$ forces $\rho$, and thus $\sigma$ to have that form.
So there are numbers $c$ and $d$ with
$$
\sigma = c\Phi^+(V)\Phi^+(V)^\dagger + d.
$$
Then $[\rho,\sigma]=0$, and it is straightforward to solve for $c$ and
$d$:
$$
\sigma = {1\over \dim V+1} \Phi^+(V)\Phi^+(V)^\dagger
       + {1\over \dim V(\dim V+1)}.
$$
In other words, $\sigma$ is the isotropic state of fidelity $1/\dim V$.
Computing $S(\rho||\sigma)$, we find:

\proclaim{Theorem 7}
Let $\rho$ be the isotropic state of fidelity $F\ge {1\over K}$ and
dimension $K$.  Then
$$
B_\Gamma(\rho) = \log_2 K + F \log_2 F + (1-F)\log_2((1-F)/(K-1)).
$$
Moreover, for all $n\ge 1$,
$$
B_\Gamma(\rho^{\otimes n}) = nB_\Gamma(\rho)
$$
\endproclaim

The second statement follows from theorem 5; we readily verify
$$
(D_\sigma \Tr(\rho\log\sigma))^\Gamma \ge -1.
$$

Remark. (1) If $\rho$ is maximally entangled, then, while there exists
an optimal $\sigma$ with full symmetry, there also exist optimal $\sigma$
with much smaller symmetry groups.  (2) Since $B_\Gamma$ is a lower bound
on entanglement of formation (by the same proof used in
\cite\VPbound\ to show that the Vedral-Plenio bound is less than
entanglement of formation), this theorem gives a lower bound on the
entanglement of formation of an isotropic state.

Example 2.  In Example 1, the symmetries of $\rho$ were enough to force
$[\rho,\sigma]=0$, and thus to force the bound to be additive with respect
to $\rho$ (modulo the derivative bound condition).  Here we use that idea
to give a large family of states for which the bound is additive.

Let $G$ be a finite abelian group.  For an element $g\in G$ and a character
$\chi:G\to {\Bbb C}$, define the ``generalized Bell state'' associated to
$g$ and $\chi$ to be
$$
v_{g,\chi} = {1\over \sqrt{|G|}} \sum_{h\in G} \overline{\chi(h)}
\ket{h,h-g}.
$$
If $G={\Bbb Z}_2$, we recover the usual Bell states.  In general, the
generalized Bell states for a given group $G$ form an orthonormal basis of
maximally entangled states.

\proclaim{Theorem 8}
Suppose $\rho$ is a mixture of generalized Bell states for the group $G$.
Then there exists $\sigma$ optimal for $\rho$ which is also a mixture
of generalized Bell states for $G$, and thus $[\rho,\sigma]=0$.
\endproclaim

\demo{Proof}
Let $V_G$ be the Hilbert space with basis $\ket{h}$ for $h\in G$.
Define operators $X(g)$ and $Z(\chi)$ by
$$
\align
X(g)\ket{h} &= \ket{h+g},\\
Z(\chi)\ket{h} &= \chi(h) \ket{h}.
\endalign
$$
Let $G$ be the group generated by $X(g)\otimes X(g)$ for all $g$ and
$Z(\chi)\otimes Z(\overline{\chi})$ for all $\chi$.  Then each
generalized Bell state is a common eigenvector for $G$, since
we compute
$$
\align
X(g')\otimes X(g') v_{g,\chi} &= \chi(g') v_{g,\chi}\\
Z(\chi')\otimes Z(\overline{\chi'}) v_{g,\chi} &= \chi'(g) v_{g,\chi}.
\endalign
$$
Moreover, for any two generalized Bell states, there exists an element
of $G$ that distinguishes them.  From this, it follows that a state
is invariant under $G$ if and only if it is a mixture of generalized
Bell states.  Since any two such mixtures commute, it follows that
$[\rho,\sigma]=0$ for some $\sigma$ optimal for $\rho$.
\qed\enddemo

For $G={\Bbb Z}_2$, we can explicitly solve for $\sigma$; if
$$
\rho = a v_{00} v_{00}^\dagger + b v_{01} v_{01}^\dagger + 
       c v_{10} v_{10}^\dagger + d v_{11} v_{11}^\dagger,
$$
with $a\ge {1\over 2} \ge b\ge c\ge d$, then
$$
\sigma = {1\over 2} v_{00} v_{00}^\dagger + 
         {b\over 2(1-a)} v_{01}v_{01}^\dagger+
         {c\over 2(1-a)} v_{10}v_{10}^\dagger+
         {d\over 2(1-a)} v_{11}v_{11}^\dagger,
$$
and
$$
B_\Gamma(\rho) = 1+a\log_2 a+(1-a)\log_2(1-a),
$$
agreeing with the bound of \cite\epp.  In general, it appears to be rather
more difficult to compute the optimal $\sigma$ analytically.  Of particular
interest would be the case ${\Bbb Z}_2^n$, corresponding to mixtures of
tensor products of the usual Bell states.

\head Maximally correlated states\endhead

We close with consideration of a class of states for which it is
a reasonable conjecture that the p.p.t. bound is not only tight,
but is in fact equal to the 1-locally distillable entanglement.
This also gives examples of states for which the bound is additive,
even though theorem 5 does not apply.

Say that a state $\rho$ on $V\otimes V$ is {\it maximally correlated} if for
any classical measurement on $V$, Alice and Bob will always obtain
the same result.  In other words, $\rho$ is of the form
$$
\rho = \sum_{i,j} \alpha_{ij} \ket{ii}\bra{jj}
$$
for some Hermitian, trace 1 operator $\alpha$ on $V$.

\proclaim{Theorem 9}
If $\rho$ is maximally correlated, then
$$
B_\Gamma(\rho) = S(\Tr_A(\rho))-S(\rho).
$$
For any other maximally correlated state $\rho'$, 
$$
B_\Gamma(\rho\otimes \rho') = B_\Gamma(\rho)+B_\Gamma(\rho').
$$
\endproclaim

\demo{Proof}
We first show that the state
$$
\sigma = \sum_i \alpha_{ii} \ket{ii}\bra{ii}
$$
is optimal for $\rho$.  Certainly, $\sigma$ is p.p.t. (indeed, it is
manifestly separable), so it remains to apply the Karush-Kuhn-Tucker
condition.  Of course, $\sigma$ is only semi-definite, so we must
use the more complicated condition following Theorem 4.

We have
$$
D_\sigma \Tr(\rho\log(\sigma)) = 
\sum_{i,j} \alpha_{ij} f(\alpha_{ii},\alpha_{jj}) \ket{ii}\bra{jj},
$$
where
$$
f(\alpha,\beta) = {\log\alpha-\log\beta\over \alpha-\beta},
$$
except that
$$
f(\alpha,\alpha) = {1\over \alpha}.
$$
We thus choose
$$
\align
L &= \sum_{i\ne j} \lambda_{ij} \ket{ij}\bra{ij}\\
K &= -\sum_{i\ne j} \alpha_{ij} f(\alpha_{ii},\alpha_{jj}) \ket{ii}\bra{jj}\\
  &\phantom{{}={}} +\sum_{i\ne j} (1-\lambda_{ij}) \ket{ij}\bra{ij},
\endalign
$$
for suitable numbers $\lambda_{ij}=\lambda_{ji}$.  We observe that
$$
\sigma L = \sigma^\Gamma K^\Gamma = 0,
$$
so it remains to verify that $L,K^\Gamma\ge 0$.  Clearly $L\ge 0$ if
and only if each $\lambda_{ij}\ge 0$, while $K^\Gamma$ is essentially a
block matrix with $2\times 2$ blocks
$$
\pmatrix
1-\lambda_{ij} & -\alpha_{ij} f(\alpha_{ii},\alpha_{jj})\\
-\alpha_{ji} f(\alpha_{ii},\alpha_{jj})&1-\lambda_{ij}
\endpmatrix.
$$
We thus obtain the conditions
$$
\align
0\le \lambda_{ij}&\le 1\\
1-\lambda_{ij} &\ge |\alpha_{ij}| f(\alpha_{ii},\alpha_{jj}).
\endalign
$$
We can choose $\lambda_{ij}$ satisfying these conditions if and only if
$$
|\alpha_{ij}| f(\alpha_{ii},\alpha_{jj})\le 1.
$$
But, since $\alpha$ is positive,
$$
(|\alpha_{ij}| f(\alpha_{ii},\alpha_{jj}))^2
\le
\alpha_{ii}\alpha_{jj} f(\alpha_{ii},\alpha_{jj})^2
=
{\beta \log(\beta)^2\over (1-\beta)^2}
\le 1,
$$
where $\beta = \alpha_{jj}/\alpha_{ii}$.

Additivity follows from the fact that the tensor product of
maximally correlated states is maximally correlated.
\qed\enddemo

In particular,

\proclaim{Corollary 2}
If $\rho$ is a pure state, then
$$
B_\Gamma(\rho) = E_f(\rho).
$$
\endproclaim

\demo{Proof}
Increasing the dimension of $V_A$ or $V_B$ as necessary to make the
dimensions equal, we find that $\rho$ is locally equivalent to a
maximally correlated state.  The result then follows immediately from
theorem 9.
\qed\enddemo

Here, in fact, $D_1(\rho)=B_\Gamma(\rho)$, by the fact \cite\puredistill\
that pure states can be distilled at a rate equal to their entanglement of
formation.  We also have

\proclaim{Corollary 3}
Let $\rho$ be a maximally correlated state on a $2\times 2$ dimensional
Hilbert space.  Suppose $\Tr_A(\rho)=1/2$.  Then
$$
D_1(\rho) = B_\Gamma(\rho) = 1-S(\rho).
$$
\endproclaim

\demo{Proof}
This is equivalent to a mixture of two Bell states.  As remarked in
\cite\epp, we can distill this using classical error correcting
codes, and find
$$
D_1(\rho)=1-S(\rho).
$$
\qed\enddemo

Since these two examples are at opposite extremes in a certain sense,
the following conjecture seems reasonable:

\proclaim{Conjecture}
For any maximally correlated state $\rho$,
$$
D_1(\rho)=B_\Gamma(\rho).
$$
\endproclaim

\head Acknowledgements\endhead

The author would like to thank J. Smolin and especially P. Shor for
helpful conversations.

\enddocument
\bye